\documentclass[aps,prl,twocolumn,superscriptaddress]{revtex4-1}
\usepackage[colorlinks=true, pdfstartview=FitV, linkcolor=red, citecolor=blue, urlcolor=black, pdftitle={},pdfauthor={Tomoya Hayata},pdfsubject={}, pdfkeywords={}]{hyperref}
\usepackage{graphicx}
\usepackage{setspace,bm}
\usepackage{latexsym,amssymb,amsmath,mathrsfs}
\usepackage{color}
\usepackage{times}
\graphicspath{{./figure/}}

\makeatletter
\providecommand \@ifxundefined [1]{%
 \@ifx{#1\undefined}
}%
\providecommand \@ifnum [1]{%
 \ifnum #1\expandafter \@firstoftwo
 \else \expandafter \@secondoftwo
 \fi
}%
\providecommand \@ifx [1]{%
 \ifx #1\expandafter \@firstoftwo
 \else \expandafter \@secondoftwo
 \fi
}%
\providecommand \enquote  [1]{``#1''}%
\providecommand \bibnamefont  [1]{#1}%
\providecommand \bibfnamefont [1]{#1}%
\providecommand \citenamefont [1]{#1}%
\providecommand \href@noop [0]{\@secondoftwo}%
\providecommand \href [0]{\begingroup \@sanitize@url \@href}%
\providecommand \@href[1]{\@@startlink{#1}\@@href}%
\providecommand \@@href[1]{\endgroup#1\@@endlink}%
\providecommand \@sanitize@url [0]{\catcode `\\12\catcode `\$12\catcode
  `\&12\catcode `\#12\catcode `\^12\catcode `\_12\catcode `\%12\relax}%
\providecommand \@@startlink[1]{}%
\providecommand \@@endlink[0]{}%
\providecommand \url  [0]{\begingroup\@sanitize@url \@url }%
\providecommand \@url [1]{\endgroup\@href {#1}{\urlprefix }}%
\providecommand \urlprefix  [0]{URL }%
\providecommand \Eprint [0]{\href }%
\providecommand \doibase [0]{http://dx.doi.org/}%
\providecommand \selectlanguage [0]{\@gobble}%
\providecommand \bibinfo  [0]{\@secondoftwo}%
\providecommand \bibfield  [0]{\@secondoftwo}%
\providecommand \BibitemOpen [0]{}%
\providecommand \EOS [0]{\spacefactor3000\relax}%
\providecommand \BibitemShut  [1]{\csname bibitem#1\endcsname}%
\let\auto@bib@innerbib\@empty
\newcommand{\be}{\begin{equation}}      
\newcommand{\ee}{\end{equation}}      
\newcommand{\bea}{\begin{eqnarray}}      
\newcommand{\eea}{\end{eqnarray}}

\begin{document}

\title{Negative refractive index in cubic noncentrosymmetric superconductors}

\author{Tomoya Hayata}
\affiliation{Department of Physics, Chuo University, 1-13-27 Kasuga, Bunkyo, Tokyo, 112-8551, Japan}
\email{hayata@phys.chuo-u.ac.jp}

\date{\today}

\begin{abstract}
We study the negative refractive index in cubic noncentrosymmetric superconductors.
We consider the Maxwell equations under the Meissner effect, and magnetoelectric effect arising due to broken inversion symmetry.
We derive dispersion relations of electromagnetic waves, and show that the refractive index becomes negative at frequencies just below the Higgs gap.
We find that the chiral mechanism of the negative refractive index, which is usually discussed in chiral materials with negative permittivity can be applied to superconductors with positive permittivity by replacing the plasma gap with the Higgs gap.
Estimation from the experimental values of the penetration depth of LiPt$_3$B indicates that the negative refractive index may appear in UV regions. 
LiPt$_3$B may exhibit the negative refractive index at wavelengths shorter than any other material observed so far.
\end{abstract}

\maketitle

{\it Introduction.} Magnetoelectric effect has caught the attention of researchers in studies of condensed matter systems with broken time-reversal or inversion symmetry~\cite{0022-3727-38-8-R01,0953-8984-20-43-434203,RevModPhys.82.3045,RevModPhys.83.1057}.
The cross polarization/magnetization is induced by magnetic/electric fields, and leads to the optical phenomena such as optical activity~\cite{Landau}, or to the transport phenomena such as 
gyrotropic current induced by AC magnetic fields in noncentrosymmetric condcutors~\cite{Levitov} ($\bm j=-\alpha\partial_t\bm B$).
Such an AC current was rediscovered in the study of the chiral magnetic effect~\cite{Fukushima:2008xe} in Weyl semimetals with broken inversion symmetry~\cite{zhong2016gyrotropic}, 
and is sometimes referred to as the gyromagnetic effect or the dynamic chiral magnetic effect.

Magnetoelectric effect plays an essential role to realize the negative refractive index in chiral materials~\cite{Tretyakov,Pendry1353,PhysRevLett.95.123904,PhysRevLett.102.023901} or topological metals~\cite{2018arXiv180110272H}.
In the original route proposed by Veselago~\cite{Veselago}, the negative refractive index is realized via double negative permittivity and permeability,
which are unlikely in natural materials, and realized only in artificial metamaterials~\cite{Smith788}. 
However in the chiral route, the negative refractive index can be realized without the unnatural assumption. 
Thus it may open a chance for broad applications.
The chiral mechanism is based on two effects: (I) The degenerate gap of transverse waves originates from the plasma screening.
(II) The helical lift of the degeneracy in momentum space originates from the magnetoelectric effect~\cite{Tretyakov,Pendry1353,PhysRevLett.95.123904,PhysRevLett.102.023901} or the chiral magnetic effect~\cite{2018arXiv180110272H}, which is analogous to the lift of the spin degeneracy due to the Rashba spin-orbit coupling~\cite{Rashba}.

Now we address a question whether other types of materials exhibit similar effects or not. 
(I) It is well known that photons acquire the mass (gap) in superconductors via the Meissner effect or the Higgs mechanism.
(II) It is known that magnetoelectric effect arises in superconductors with broken inversion symmetry~\cite{Agterberg2012,Fujimoto2012}.
Then we expect that noncentrosymmetric superconductors exhibit the negative refractive index.

In this Letter, we discuss the negative refractive index in cubic noncentrosymmetric superconductors.
We study the Maxwell equations under the Meissner and magnetoelectric effects. 
By computing dispersion relations of electromagnetic waves, we show that 
the refractive index indeed becomes negative for either of circular polarizations at frequencies below the Higgs gap.
We estimate the Higgs gap from the penetration depth observed in experiments, which indicates that the negative refractive index may occur in UV regions. 
Transmission coefficients in an insulator-superconductor-insulator junction are computed as an experimental implication.

{\it Magnetoelectric effect in cubic noncentrosymmetric superconductors.} 
We start with a computation of electric current in cubic noncentrosymmetric superconductors under electromagnetic fields based on the Ginzburg-Landau theory. 
In noncentrosymmetric superconductors, electric current is induced by magnetic fields via  magnetoelectric effect, in addition to the standard diamagnetic current. 
The  Ginzburg-Landau free energy for the order parameter field $\psi$ is given as~\cite{Agterberg2012,Fujimoto2012}
\bea
F&=& \int d^3x\;f_{\rm n}+f_{\rm B}+f_{\rm s}+f_{\rm LI} ,
\label{eq:GLfree}
\\
f_{\rm B}&=& \frac{1}{8\pi}\bm B^2 ,
\\
f_{\rm s}&=& \frac{1}{2m^*}\left|\left(\frac{\hbar}{i}\nabla+\frac{e^*}{c} \bm A\right)\psi\right|^2-\alpha |\psi|^2+\frac{\beta}{2}|\psi|^4 ,
\\
f_{\rm LI}&=& \frac{1}{2}\sum_{i,j}d_{ij} B_i\cdot{\rm Re}\left[\psi^*\left(\frac{\hbar}{i}\nabla+\frac{e^*}{c} \bm A\right)_j\psi\right] ,
\label{eq:LIfree}
\eea
where $m^*$, and $-e^*=-2e$ are the mass and electric charge of the Cooper pair ($e>0$ is the elementary charge).
$\hbar$, and $c$ are the plank constant, and speed of light. 
$\bm A$, and  $\bm B=\nabla\times\bm A$ are the vector potentials, and magnetic fields. 
$F_{\rm n}=\int d^3xf_{\rm n}$ is the free energy of the normal state and independent of $\psi$. 
$F_{\rm B}=\int d^3xf_{\rm B}$ is the magnetic energy. 
$F_{\rm s}=\int d^3xf_{\rm s}$ is the conventional free energy of the super component.
$F_{\rm LI}=\int d^3xf_{\rm LI}$ is the so called Lifshitz invariant. 
It is invariant under time-reversal operation, but not under spatial-inversion operation, so that it characterizes the noncentrosymmetric nature of superconductors. 
Below we consider a cubic system, and take $d_{ij}=d \delta_{ij}$.

From variation of Eq.~\eqref{eq:GLfree} with respect to $\bm A$, we obtain
\bea
&&\frac{c}{4\pi}\nabla\times\bm B 
\notag
\\
&&=\bm j_{\rm n}
-\frac{e^*}{m^*}\left(\hbar\nabla\theta+\frac{e^*}{c}\bm A\right)|\psi|^2
\notag
\\
&&-d e^*\bm B|\psi|^2
-dc\nabla\times\left(\left(\hbar\nabla\theta+\frac{e^*}{c}\bm A\right)|\psi|^2\right) ,
\label{eq:Maxwell}
\eea
where $\psi=|\psi|e^{i\theta}$. The sum of four terms in the right hand side is the total electric current. 
The first term $\bm j_{\rm n}=\delta F_{\rm n}/\delta \bm A$ is the normal current. 
It vanishes in equilibrium.
The second term $\bm j_{\rm s}=-\frac{e^*}{m^*}\left(\hbar\nabla\theta+\frac{e^*}{c}\bm A\right)|\psi|^2$ is the standard superconducting current.
The last two terms express the effect of the inversion symmetry breaking. 
In a uniform $\psi$, these terms can be understood as the polarization current $\partial_t\bm P$ and the magnetization current $c\nabla \times \bm M$ induced by the magnetoelectric effect.
Let us first discuss the magnetization current.
From variation of $F_{\rm LI}$ with respect to $\bm B$, the magnetization from the supercomponent $\bm M_{\rm s}$ is given as
\bea
\bm M_{\rm s} =-\frac{\delta F_{\rm LI}}{\delta \bm B}=
-d\left(\hbar\nabla\theta+\frac{e^*}{c}\bm A\right)|\psi|^2 .
\label{eq:magnetization}
\eea
Therefore the last term in Eq.~\eqref{eq:Maxwell} is nothing but the magnetization current $c\nabla\times\bm M_{\rm s}$.
This part does not require any assumption on $\psi$.
Next let us discuss the polarization current.
In terms of the magnetoelectric effect, it is useful to rewrite the Lifshitz invariant~\eqref{eq:LIfree} as
\bea
F_{\rm LI}
=\int d^3x\;&&\frac{cd}{i\omega}\bm E\cdot \left(\nabla\times\left(\hbar\nabla\theta+\frac{e^*}{c} \bm A\right) |\psi|^2\right)
\notag
\\
+&&\frac{cd}{i\omega}\nabla\cdot\left(\bm E\times\left(\hbar\nabla\theta+\frac{e^*}{c} \bm A\right) |\psi|^2\right) ,
\label{eq:LI}
\eea
where we assumed a monochromatic $\bm A$, and used the Bianchi identity: $\nabla\times\bm E=-\partial_t\bm B/c=i\omega\bm B/c$.
The first term in Eq.~\eqref{eq:LI} indicates the bulk polarization from inhomogeneous supercurrent:
\bea
\bm P_{\rm s} =\frac{\delta F_{\rm LI}}{\delta \bm E}
=\frac{cd}{i\omega}\nabla\times\left(\left(\hbar\nabla\theta+\frac{e^*}{c} \bm A\right) |\psi|^2\right) .
\label{eq:polarization}
\eea
The second term in Eq.~\eqref{eq:LI} indicates the surface Hall conduction, which is elucidated below.
Equation~\eqref{eq:polarization} is valid in general cases.
In particular, if the density of the order parameter is uniform ($|\psi|^2=n_s=\alpha/\beta$) [We assume that $\alpha$, $\beta>0$], 
and no vortex exists ($\nabla\times\nabla\theta=0$), it reads
\bea
\bm P_{\rm s} 
=\frac{de^*n_s}{i\omega}\bm B .
\eea
Thus the polarization current $\partial_t\bm P_{\rm s}$ equals to the third term in Eq.~\eqref{eq:Maxwell}.
In fact, if $|\psi|^2=n_s$, and $\nabla\times\nabla\theta=0$, the bulk term in Eq.~\eqref{eq:LI} is rewritten as 
\bea
\tilde{F}_{\rm LI}
=\int d^3x\;\frac{de^*n_s}{i\omega}\bm E\cdot\bm B ,
\label{eq:LI3}
\eea
and we obtain the cross polarization and magnetization:
\bea
\bm P_{\rm s} 
&=&\frac{\delta \tilde{F}_{\rm LI}}{\delta \bm E}=
\frac{de^*n_s}{i\omega}\bm B ,
\label{eq:polarization2}
\\
\bm M_{\rm s} 
&=&-\frac{\delta \tilde{F}_{\rm LI}}{\delta \bm B}=-\frac{de^*n_s}{i\omega}\bm E .
\label{eq:magnetization2}
\eea
Equation~\eqref{eq:LI3} belongs to the pseudoscalar in the general classification of the linear magnetoelectric effects~\cite{0022-3727-38-8-R01,0953-8984-20-43-434203}. 
The polarization and magnetization in Eqs.~\eqref{eq:polarization2} and~\eqref{eq:magnetization2} have the same form with the clean Weyl semimetals with broken inversion symmetry~\cite{zhong2016gyrotropic}.
Although the gyromagnetic effect in noncentrosymmetric conductors induces no DC current ($\bm j\sim \partial_t\bm B$),
it in noncentrosymmetric superconductors survives even in static $\bm B$ ($\bm j\sim \bm B$).
This is because the polarization and magnetization currents in Eq.~\eqref{eq:Maxwell} originate from the inhomogeneous supercurrent $\nabla\times\bm j_{\rm s}\sim n_s\bm B$. 
This static magnetoelectric effect may be important characteristic of noncentrosymmetric superconductors~\cite{Agterberg2012,Fujimoto2012}.

Below we consider the uniform state, in which $|\psi|^2=n_s$, and $\nabla\theta=0$ by a gauge choice. 
Then we have 
\bea
\bm j_{\rm tot}
=-\frac{(e^*)^2n_s}{m^*c}\bm A-2dn_se^*\bm B .
\label{eq:tot}
\eea

{\it Negative refractive index.} 
We study the propagation of monochromatic electromagnetic waves. 
With the polarization~\eqref{eq:polarization2}, and electric current~\eqref{eq:tot}, the Maxwell equations are
\bea
&&\nabla\cdot\left(\epsilon\bm E+\frac{de^*n_s}{i\omega}\bm B \right)=0 ,
\label{eq:Maxwell1}\\
&&\nabla \cdot\bm B=0 ,
\label{eq:Maxwell2}\\
&&\nabla\times \bm E =-\frac{1}{c}\partial_t \bm B ,
\label{eq:Maxwell3}\\
&&\nabla\times \bm B =\frac{1}{c}\partial_t\epsilon \bm E+\frac{4\pi}{c} \bm j_{\rm tot},
\label{eq:Maxwell4}
\eea
where $\bm E=\tilde{\bm E}e^{i\bm p\cdot\bm x-i\omega t}$, and $\bm B=\tilde{\bm B}e^{i\bm p\cdot\bm x-i\omega t}$ are electric, and magnetic fields, respectively. 
Below we assume that the permittivity $\epsilon$ is isotropic and unity for simplicity.
Then the Maxwell equations are rewritten as
\bea
\left(\frac{\omega^2}{c^2}-\bm p^2-\frac{1}{\lambda^2}-\frac{2\delta}{\lambda^2} \bm t\cdot\bm p \right)\bm E &=&0,
\label{eq:Maxwell5}
\\
\left(\frac{\omega^2}{c^2}-\bm p^2-\frac{1}{\lambda^2}-\frac{2\delta}{\lambda^2} \bm t\cdot\bm p \right)\bm B &=&0 ,
\label{eq:Maxwell6}
\eea
where $\lambda=\sqrt{m^*c^2/(4\pi n_s(e^*)^2)}$, and $\delta=m^*c d/e^*$ are the penetration depth, and the effective length characterizing the inversion-symmetry breaking, and $i\bm p\times\bm E= (\bm t\cdot\bm p)\bm E$, with $[t_i]_{jk}=-i\epsilon_{ijk}$. 
The Maxwell equations~\eqref{eq:Maxwell5}, and~\eqref{eq:Maxwell6} are diagonal in circular polarizations satisfying
\bea
i\bm p\times\bm E_{\pm}= \bm t\cdot\bm p\;\bm E_{\pm}=  \pm p\bm E_{\pm} , \\
i\bm p\times\bm B_{\pm}=\bm t\cdot\bm p\;\bm B_{\pm}=  \pm p\bm B_{\pm} .
\eea
The dispersion relations read
\bea
\omega_{\chi\pm}=\pm c \sqrt{\left(p_\chi+ \chi\frac{\delta}{\lambda^2}\right)^2+\frac{1}{\tilde{\lambda}^2}} ,
\label{eq:dispersion}
\eea
for the right-handed ($\chi=+$) [left-handed ($\chi=-$)] circular polarization, and $\tilde{\lambda}=\lambda/\sqrt{1-\delta^2/\lambda^2}$ is the effective penetration depth~\cite{PhysRevB.77.054515}.
The dispersion relations~\eqref{eq:dispersion} show the same behavior with Ref.~\cite{2018arXiv180110272H}, so that the analysis in Ref.~\cite{2018arXiv180110272H} can directly be applied if we replace the plasma gap $\omega_p$ by the higgs gap $c/\lambda$.
As will be discussed in detail, the dispersion relations~\eqref{eq:dispersion} show the different characteristic behaviors in three regimes: ($a$) $\delta=0$, ($b$) $|\delta|<\lambda$, and ($c$) $|\delta|>\lambda$.
When the system is centrosymmetric, that is, $\delta=0$, two transverse modes are degenerate with the gapped dispersion relations. The electromagnetic fields are exponentially expelled from superconductors, which is known as the Meissner effect.
In noncentrosymmetric superconductors with nonzero $\delta$, the helical degeneracy is lifted, and either of them has the negative refractive index just below the higgs gap $c/\lambda$.

Now we discuss the negative refractive index. 
We first study the negative sign of the group velocity $v_{g\chi}$ compared to the phase velocity $v_{p\chi}$~\cite{Pendry1353,2018arXiv180110272H}.
These are given as
\bea
\frac{v_{g\chi}}{c}&=&\frac{\partial \omega_{\chi+}}{\partial cp_\chi}
=\frac{p_\chi+\chi\frac{\delta}{\lambda^2}}{\sqrt{\left(p_\chi+\chi\frac{\delta}{\lambda}\right)^2+\frac{1}{\tilde{\lambda}^2}}} ,
\label{eq:group}
\\
\frac{v_{p\chi}}{c}&=&\frac{\omega_{\chi+}}{cp_\chi}
=\frac{\sqrt{\left(p_\chi+\chi\frac{\delta}{\lambda}\right)^2+\frac{1}{\tilde{\lambda}^2}}}{p_\chi} .
\eea
We consider the regime ($b$) $|\delta|<\lambda$.
Then the argument of the square root is positive. 
$v_{g\chi}$ has the negative sign compared to $v_{p\chi}$ 
when $|p_\chi|<|\delta|/\lambda^2$, that is, when $c/\tilde{\lambda}<\omega_\chi<c/\lambda$ for either of circular polarizations.
If $\omega_\chi<c/\tilde{\lambda}$, both of them are expelled from superconductors, so that $\tilde{\lambda}$ works as the effective penetration depth.
Next we consider the regime ($c$) $|\delta|>\lambda$. 
Then the dispersion relations~\eqref{eq:dispersion} become pure imaginary at some momentum range and indicate unstable modes analogous to chiral plasma instabilities~\cite{PhysRevLett.111.052002}. 
When we discuss transmission of electromagnetic waves,
we need only the refractive index, that is, $n(\omega)=cp(\omega)/\omega$ with positive $\omega$ as information of dispersion relations, so that we do not encounter the unstable modes.
It would be interesting to study real-time dynamics with parameters in the regime ($c$),
where an analogue of chiral plasma instabilities may affect the dynamic transition between the magnetic helicity and (quantized) vorticity of quantum vorticies.
Anyway, we consider positive $\omega$. Then gap appears at 
$|p_\chi+\chi \delta/\lambda^2|<\sqrt{(\delta^2-\lambda^2)}/\lambda^2$. 
Below the gap, the argument is still positive, and $v_{g\chi}$ has the negative sign compared to $v_{p\chi}$ at all frequencies lower than $c/\lambda$.

We can directly discuss the negative refractive index. By rewriting Eq.~\eqref{eq:dispersion} into the $\omega$-dependence of $p$,
the refractive index reads
\bea
n_{\chi}=\frac{cp_{\chi+}}{\omega_\chi}
=\frac{-\chi\frac{c\delta}{\lambda^2}+\sqrt{\omega_\chi^2-\frac{c^2}{\tilde{\lambda}^2}}}{\omega_\chi} ,
\label{eq:refractive}
\eea
from the positive sign solution of the quadratic equation on $p$.
In the regime ($b$) $|\delta|<\lambda$, we have $0<\sqrt{\omega_\chi^2-c^2/\tilde{\lambda}^2}<c|\delta|/\lambda^2$ at $c/\tilde{\lambda}<\omega_\chi<c/\lambda$, so that the refractive index~\eqref{eq:refractive} is negative for either of circular polarizations. 
Next in the regime ($c$) $|\delta|>\lambda$, 
we have $c\sqrt{\delta^2-\lambda^2}/\lambda<\sqrt{\omega_\chi^2-c^2/\tilde{\lambda}^2}<c|\delta|/\lambda^2$ at $0<\omega_\chi<c/\lambda$,
so that the refractive index~\eqref{eq:refractive} is negative for either of circular polarizations.
The Higgs gap vanishes in the regime ($c$), which means that electromagnetic waves can propagate inside superconductors even though they experience the Meissner effect. 
This can be understood from the fact that the effective penetration depth $\tilde{\lambda}$ becomes pure imaginary, and can explicitly be seen from the refractive index~\eqref{eq:refractive}.
In the regime ($a$) or ($b$), the refractive index has an imaginary part if $\omega_\chi<c/\tilde{\lambda}$, and electromagnetic waves become evanescent. 
On the other hand, the argument of the square root in Eq.~\eqref{eq:refractive} is always positive in the regime ($c$).
No evanescent wave appears, and even the standing wave with $\omega_\chi=0$ is possible although it is spatially-inhomogeneous~\cite{PhysRevB.77.054515}.

In this analysis, we consider only the Lifshitz invariant~\eqref{eq:LIfree} as the effect of the inversion symmetry breaking~\cite{Agterberg2012,Fujimoto2012}.
This may implicitly assume that $\delta$ is small compared to $\lambda$.
In practical applications, we may need to take higher order terms of the inversion symmetry breaking into account to study the regime ($c$).
This can be done e.g., by computing the electric current~\eqref{eq:tot} from the microscopic theory based on the Kubo formula.
We note that in the first order of $\delta/\lambda$, the negative refractive index disappears since $\tilde{\lambda}=\lambda$.

The representative materials of cubic noncentrosymmetric superconductors are Li(Pd$_{1-x}$Pt$_x$)$_3$B (We also need the absence of mirror symmetry to have nonvanishing magnetoelectirc effect)~\cite{Agterberg2012,Fujimoto2012}.
The penetration depth of these materials is $190$-$364$nm at zero temperature~\cite{JPSJ.74.1014}, and is less than $400$nm at low temperatures~\cite{PhysRevLett.97.017006},
so that the negative refractive index may occur in UV regions, that is, wavelengths shorter than any other material observed so far.
Among them, LiPt$_3$B is expected to have the large inversion symmetry breaking~\cite{PhysRevLett.97.017006}, 
and is most likely to exhibit the negative refractive index.

\begin{figure}[t]
\centering
 \includegraphics[width=.4\textwidth]{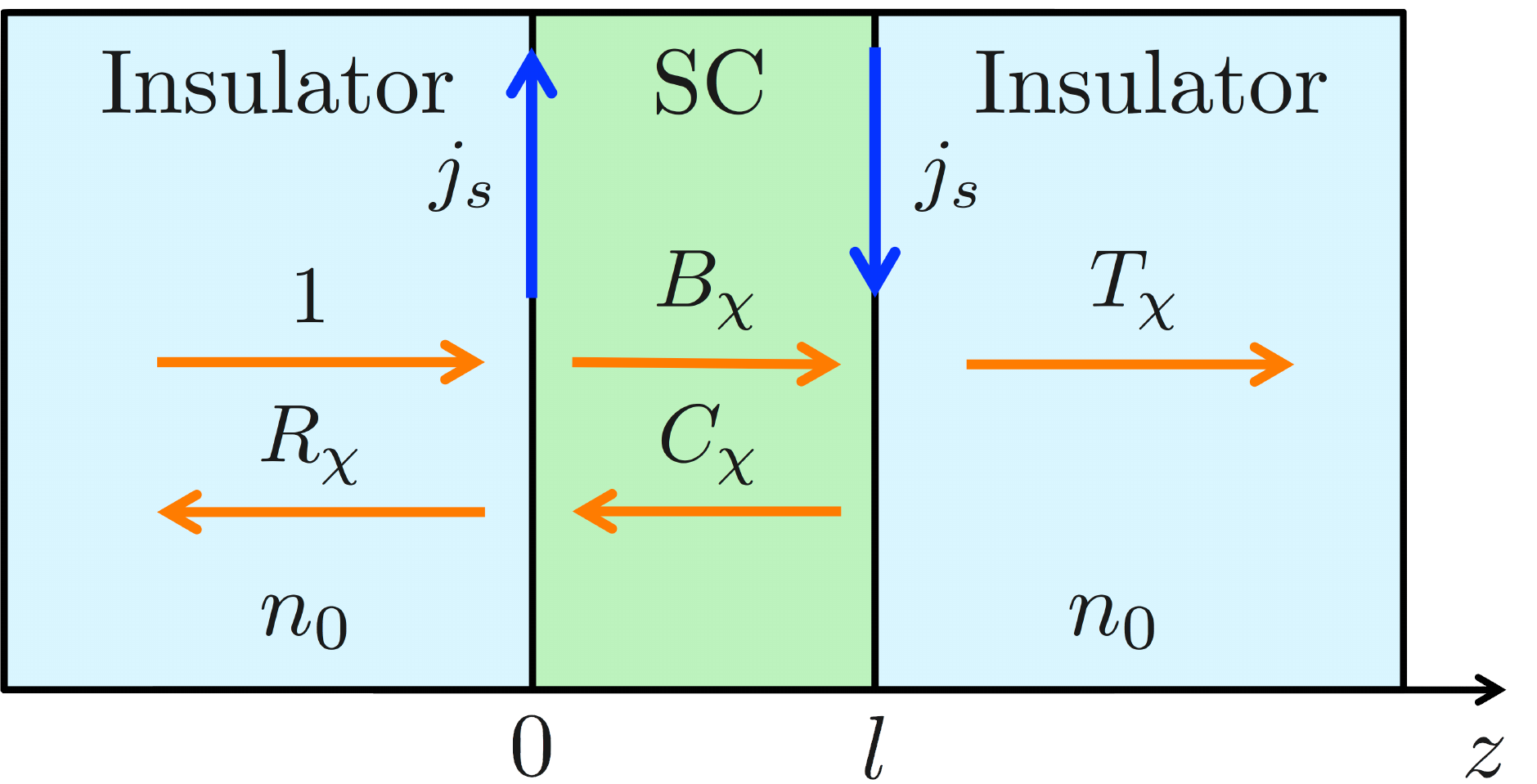}
 \caption{
Incident, reflected, and transmitted electric waves in insulator-superconductor (SC)-insulator junction. 
The surface Hall conduction is induced by the incident waves.
\label{fig2}
}
\end{figure}

{\it Transmission coefficients.} 
We here compute reflection and transmission coefficients of electromagnetic waves in an insulator-superconductor-insulator junction (Thickness of superconductors is $l$).
We consider the right- or left-handed electromagnetic wave normally incident from a nonmagnetic insulator (dielectric) with the refractive index $n_0(>0)$ as shown in Fig.~\ref{fig2}.
The transmission coefficients may be used to experimentally retrieve the negative refractive index~\cite{PhysRevB.65.195104}.

When $|\psi|^2$ changes by the step function, the surface Hall conduction arises from the magnetization current in Eq.~\eqref{eq:Maxwell}: 
\bea
\bm j_s=\pm \frac{icde^*n_s}{\omega}\hat{z}\times\bm E\delta(z-z_{\rm s}) ,
\label{eq:surfaec}
\eea
where the sign is $+$ [$-$] for the left surface ($z_s=0$) [right surface ($z_s=l$)].
Due to the surface Hall conduction~\eqref{eq:surfaec}, the boundary condition on the tangential components of magnetic fields is modified as 
$\hat{z}\times\left(\bm B_{\rm Air}-\bm B_{\rm SM}\right)=-\hat{z}\times 4\pi ide^*n_s\bm E/\omega$.
The tangential components of $\tilde{\bm B}=\bm B-4\pi ide^*n_s\bm E/\omega$ are continuous at boundaries as well as $\bm E$.

The amplitude of electric fields $\bm E_\chi=E_\chi\left(\hat{x}+i\chi\hat{y}\right)$ are
\bea
E_\chi=
\begin{cases}
e^{in_0\omega z/c}-R_\chi e^{-in_0\omega z/c} & \text{$z\leq0$} \\
B_\chi e^{in_\chi \omega z/c}-C_\chi e^{-in_{-\chi}\omega z/c} & \text{$0\leq z\leq l$}\\
T_\chi e^{in_0\omega(z-l)/c} & \text{$l\leq z$}
\end{cases} ,
\label{eq:amplitude}
\eea
where we take the amplitude of the incident wave unity, and $\omega$ is frequency in an incident dielectric.
The amplitudes of magnetic waves are obtained from $\bm B_\chi=c\bm p\times\bm E_\chi/\omega$. 
Then from the continuity conditions at $z=0$, and $z=l$ on $\bm E$, and $\tilde{\bm B}$, reflection and transmission coefficients read
\bea
R_{\chi}&=&\frac{(n_r^2-1)\sin \kappa n_r}{2in_r\cos\kappa n_r+(1+n_r^2)\sin \kappa n_r} ,
\label{eq:reflection}
\\
T_\chi&=&\frac{2in_r e^{-\chi i\delta l/(n_0\lambda^2)}}{2in_r\cos\kappa n_r+(1+n_r^2)\sin \kappa n_r} ,
\label{eq:transmission}
\eea
where $n_r=(n_++n_-)/(2n_0)=\sqrt{1-c^2/(\tilde{\lambda}\omega)^2}/n_0$, and $\kappa=n_0\omega l/c$.
Eqs.~\eqref{eq:reflection}, and~\eqref{eq:transmission} reproduce those in gyrotopic medium~\cite{PhysRevB.73.045114}. 
The azimuth rotation, and ellipticity of transmitted waves are obtained from Eq.~\eqref{eq:transmission} as $\theta_{\rm T}=({\rm Arg}\;T_+-{\rm Arg}\;T_-)/2=-\delta l/n_0\lambda^2$, and $\eta_{\rm T}=(1/2)\sin^{-1}(|T_+|^2-|T_-|^2)/(|T_+|^2+|T_-|^2)=0$.
The direction of azimuth rotation is determined by the sign of $d$. 
$|T_+|$ or $|T_-|$ shows the resonant perfect transmission if $\kappa n_r=\pi m$ with $m$ being some integers, which is the same as conventional metals.
Similarly, the azimuth rotation, and ellipticity of reflected waves read $\theta_{\rm R}=\eta_{\rm R}=0$ as expected from time-reversal symmetry~\cite{PhysRevB.73.045114}.

{\it Conclusion.} 
In this Letter, we have discussed the negative refractive index in cubic noncentrosymmetric superconductors.
We have studied the Maxwell equations under the Meissner and magnetoelectric effects. 
The latter can arise due to broken inversion symmetry.
We show that the chiral mechanism of the negative refractive index~\cite{Tretyakov,Pendry1353,PhysRevLett.95.123904,PhysRevLett.102.023901,2018arXiv180110272H} can be applied to noncentrosymmetric superconductors by replacing the plasma gap with the Higgs gap.
We have also computed transmission coefficients in an insulator-superconductor-insulator junction as an experimental implication.

The candidate material to experimentally test our mechanism is Li(Pd$_{1-x}$Pt$_x$)$_3$B.
In particular, the large inversion-symmetry breaking is expected in LiPt$_3$B. 
Estimation from experimental values of the penetration depth indicates that  LiPt$_3$B may exhibit the negative refraction in UV regions.

\begin{acknowledgments}
The author thanks Y.~Hidaka for useful comments.
This work was supported by JSPS Grant-in-Aid for Scientific Research (No: JP16J02240).

\end{acknowledgments}

\bibliography{./refractive}

\end{document}